# *Healing Spaces*: Feasibility of a Multisensory Experience for Older Adults with Advanced Dementia and their Caregivers

Gabriela Purri R. Gomes
Interactive Media & Games Division
Creative Media & Behavioral Health Center
School of Cinematic Arts
University of Southern California
Los Angeles, CA USA
gomesg@usc.edu

Sydney Rubin
Interactive Media & Games Division
Creative Media & Behavioral Health Center
School of Cinematic Arts
University of Southern California
Los Angeles, CA USA
sydneyru@usc.edu

Leah I. Stein Duker
Chan Division of Occupational Science & Occupational Therapy
University of Southern California
Los Angeles, CA USA
lstein@chan.usc.edu

Donna Benton
Davis School of Gerontology
University of Southern California
Los Angeles, CA USA
benton@usc.edu

Andreas Kratky
Interactive Media & Games Division
School of Cinematic Arts
Los Angeles, CA USA
akratky@cinema.usc.edu

Sze Yu A Chen
Interactive Media & Games Division
School of Cinematic Arts
Los Angeles, CA USA
szeyuche@usc.edu

Maryalice Jordan-Marsh
Creative Media & Behavioral Health Center
School of Cinematic Arts
University of Southern California
Los Angeles, CA USA
jordanma@usc.edu

Marientina Gotsis
Creative Media & Behavioral Health Center
School of Cinematic Arts
University of Southern California
Los Angeles, CA USA
gotsis@usc.edu

## ABSTRACT

*Healing Spaces* proposes a new approach to multisensory interventions that show potential in ameliorating the behavioral and psychological symptoms of advanced dementia in older adults. Using smart technology, the project combines both digital and physical components to transform spaces and create unified, curated sensory experiences that provide meaningful context for interaction, and are easy for caregivers to deliver. A usability study was conducted for the *Healing Spaces* app followed by a feasibility evaluation of the full experience in a memory care facility recruiting caregivers, and residents in advanced stages of dementia. The feasibility evaluation successfully illuminated strengths as well as areas for improvement for the *Healing Spaces* experience in a memory care setting with older adults with advanced dementia. Caregivers and facility managers expressed interest in continuing to use Healing Spaces with the residents of the facility. Lessons learned about the technical and logistical implementation of *Healing Spaces* are discussed, as well as future directions for study design and potential therapeutic value of the experience.

## 1   Introduction

Dementia is an umbrella term for a group of symptoms that commonly include problems with memory, thinking, problem-solving, language and perception [2]. These symptoms are caused by different diseases that affect the brain, including Alzheimer's disease, which is the most common and most studied form of dementia [1]. In those affected, there is a progressive loss of nerve cells, and the symptoms gradually affect one's ability to carry out daily tasks and take part in everyday activities [2].

Advanced dementia is characterized by severe cognitive impairment and memory loss, and progressive deterioration of verbal communication [12] [4]. As individuals with advanced dementia struggle to engage in cognitive-based or social activities, they may appear to be completely unreachable and isolated, making it challenging for caregivers to engage and interact with them [10]. Foley [12] argues, however, that "within this complex illness, there are also moments of lucidity in which the person demonstrates their ability to reflect, remember, and socially engage". As the need for non-verbal communication increases due to cognitive decline, it is essential that individuals with advanced dementia are presented with alternative opportunities to express themselves and to interact meaningfully with the world and the people around them [12] [4] [15].

Older adults living with advanced dementia are especially vulnerable to sensory imbalance, as progressive neuronal losses affect their ability to process sensory stimuli [3]. Research shows that both sensory deprivation and overstimulation often result in similar behavioral and psychological symptoms exhibited in those suffering from dementia, including agitation and depression [19]. It is also ethically and scientifically important to provide care that either calms or stimulates these patients, based on their needs, as part of the continuum of providing compassionate care [26]. It is therefore crucial that those living with advanced dementia have access to balanced, meaningful sensory stimulation that adaptable to their cognitive level [3] [27].

It has been suggested that environment design has the potential to influence healing by facilitating participation in health-inducing behaviors, producing positive emotional states (e.g., happiness) and relaxation, and enabling a sense of self-efficacy [9]. This may be specifically helpful for individuals with dementia who do not have the ability or opportunity to go outside to engage in real world multisensory stimulation activities. One such environmental design is multisensory environments (MSEs), previously referred to as Snoezelen, which are spaces designed to gently stimulate the senses and improve mood and behavior [14]. They were developed in the early 1980s and were traditionally used for leisure activities involving adults with learning disabilities [14]; however, in recent years there has been a shift to utilizing MSEs as a therapeutic tool [16]. In these spaces, visual, auditory, olfactory, vestibular and/or tactile stimulation can be offered using a variety of equipment individuals can engage with, such as colored lights, optic fibers, bubble columns and solar projectors. Although research in this area is limited, studies and anecdotal evidence suggest that these environments may be an appropriate intervention for managing the behavioral and psychological symptoms of dementia, especially in the later stages where there is severe cognitive impairment [3].

The experience of the end stages of a disease like Alzheimer's does not only affect the patient but may also leave a lasting legacy on those around the individual, including family, friends, and clinicians. Therefore, caring for these diseases at end-stages must conform to end-stage ethics [28] that are palliative in nature, dignity-preserving, and with the potential of positive memory-making in the relationship between caregiver and sufferer. This last point is important for society because it can help mitigate trauma and compassion fatigue [5], which in turn have the potential to be cost-saving despite the complexity of capturing these costs [17].

While most commercial mHealth applications for this population focus on providing assistance to caregivers via delivering information about caregiving and disease management [11], we present an experience-based mHealth application that embraces the social aspect of both patients and caregivers' lives. In addition, *Healing Spaces* utilizes user-centered design and participatory design processes, which have been urged for improving the quality and validity of these apps [7] [13].

In this paper, we discuss the design process and formative research of *Healing Spaces*, an immersive, sensory experience that seeks to encourage social connection and engagement between individuals living with advanced dementia and their caregivers.

## 2 Project Background

*Healing Spaces* was developed as part of the first author's Master of Fine Arts thesis work for the University of Southern California's (USC) Interactive Media & Games Division [24]. Motivated by her own struggle to interact with her late grandmother who suffered from Alzheimer's disease, her goal was to create a tool to help older adults living with advanced dementia and their caregivers reconnect, even if just for a brief moment.

The first author sought to understand the health and behavioral challenges associated with the later stages of dementia through a different lens, based on interdisciplinary theories and influenced by her own practice in the realm of user experience design. By integrating smart technology with elements of physical play and storytelling, *Healing Spaces* proposes a new approach to therapeutic experiences that not only soothe and engage all the senses, but also encourage meaningful, embodied moments of connection.

## 3 Project Overview

This project involved a team of USC students from the Interactive Media & Games Division, and advisors who are professionals in the realms of gerontology, occupational science and therapy, nursing, and games and media design.

Inspired by the research legacy of multisensory environments and the first author's interest in healing environments and interactive technologies, *Healing Spaces* focuses on the creation of immersive, hybrid sensory experiences that leverage existing knowledge in dementia care to create restorative, demand-free spaces where individuals living with advanced dementia and their caregivers are able to engage and relax.

With *Healing Spaces*, the design and development team attempted to create a sense of place through sight, touch, smell, and sound using both physical and digital components. The goal was to "transport" users to locations that could evoke the sense of mindfulness and peace often found in nature. Curated experiences for different themes (e.g., forest, seaside) were designed using both physical props and digital content, in order to "capture" and "project" the spatial and sensory qualities of these themed environments into existing physical spaces. In each experience, immersion was attained by infusing storytelling in every detail possible, including how a space feels through light and color, as well as through the smells and textures.

To maximize *Healing Spaces'* future potential to change health behavior and to ensure the active involvement of potential users and stakeholders in the design process, the team followed a participatory design (PD) approach in the creation of the project. In the context of human-computer interaction (HCI), PD values both the empowerment and active participation of users in the design process [6]. The following strategies and techniques were utilized by the team: (a) focused observation of caregivers' day-to-day activities in the sensory room of a memory care facility prior to development and design; (b) creation of visual and functional prototypes to facilitate discussions, brainstorming sessions, and synthesis of new ideas between the student team, caregivers, facility managers and project advisors; (c) story collecting through interviews with caregivers, facility managers, and project advisors; (d) inclusion of the perspective of individuals living with advanced dementia in the early stages of design through a feasibility evaluation.

Including the end-user's perspectives was particularly important for the developers of *Healing Spaces*. Participatory design with individuals who are non-verbal or have compromised cognition calls for a new collaborative approach to co-creation that "embraces "experience" and "meaning making" as legitimate sources of knowledge" [12], the so-called "third-space" of HCI design. Muller [20] describes this collaborative space as "a fertile environment in which participants can combine diverse knowledges into new insights and plans for action", and further emphasizes its potential in ensuring "mutual learning and reciprocal validation of diverse perspectives". Most importantly, this "third-space" calls for additional and novel participatory design techniques to foster meaningful collaboration between developers and vulnerable groups [6]. By infusing *Healing Spaces* with playfulness and opportunities for reciprocal and unexpected interactions to arise, the objective was to ensure the inclusion of older adults with advanced dementia as active contributors to the development of the application.

## 3.1 One app, a connected space

An iOS app was built for the iPad to work together with Philips® Hue smart lights, allowing caregivers to transform spaces through light, colors, sounds and visuals. The goal was to extend the digital experience into the physical space (see Figures 1 & 2).

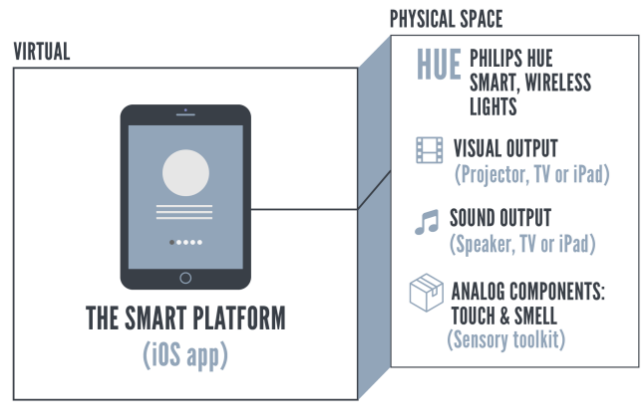

**Figure 1: Healing Spaces components. Image credits: Gabriela Purri R. Gomes.**

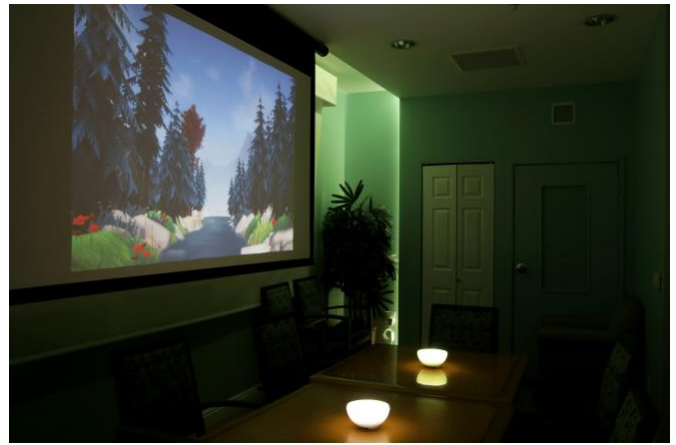

**Figure 2: Forest environment brought to life in the physical space. Image credits: Gabriela Purri R. Gomes.**

## 3.2 Setting the ambiance

The prototype built during the first author's thesis year enabled caregivers to choose between two natural settings: forest and seaside. For each, they were also able to choose a time of day: dawn (for stimulation) or dusk (for relaxation) (see Figures 3 & 4).

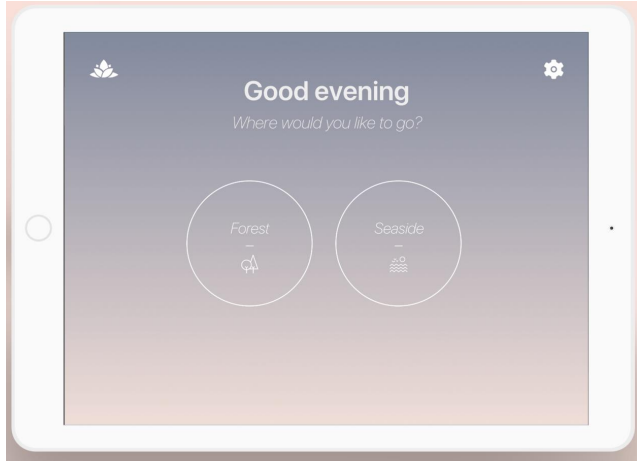

**Figure 3:** *Healing Spaces* app main screen. Image credits: Gabriela Purri R. Gomes.

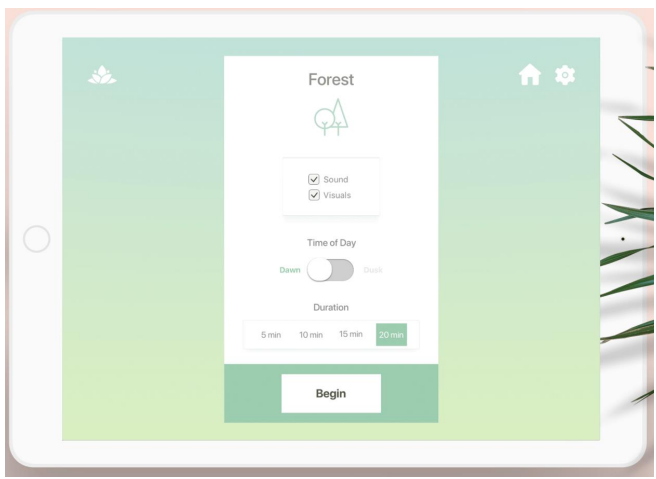

**Figure 4:** 'Settings' screen where caregivers can customize the experience. Image credits: Gabriela Purri R. Gomes.

With the tap of a button, the chosen ecosystem can be brought to life through color, natural soundscapes or music, and peaceful visuals (see Figure 2).

### 3.3 Sensory boxes

To achieve physical and emotional engagement, each digital experience was paired with a themed sensory box that included tactile, visual and olfactory sensory elements (see Figure 5). For example, the forest sensory box contained artificial moss, leaves, miniature birds and eucalyptus-scented hand lotion. Alternatively, the seaside box contained kinetic sand, sea shells and a hand lotion that smelled like sunscreen.

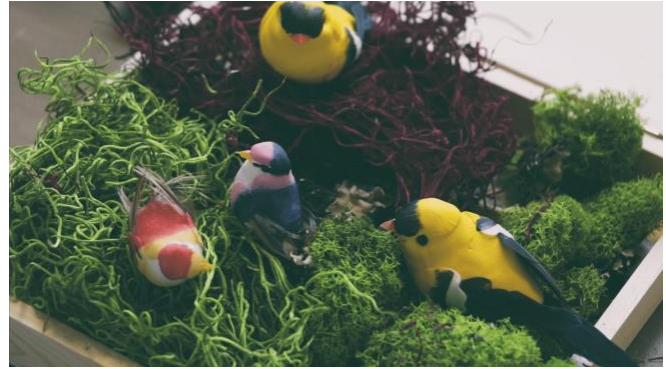

**Figure 5:** Forest sensory box. Image credits: Gabriela Purri R. Gomes.

### 3.4 Accessibility and scalability

With the goal of creating an accessible and scalable package, the team chose to use commercially available off-the-shelf physical components; reusability of each component for other aspects of care was also considered. The digital platform was also chosen with this goal in mind, as the iPad would give caregivers and end users portable access to not only the *Healing Spaces* app, but also a larger ecosystem of applications and resources. An app also offers flexibility for implementation in both home and community settings with either low- or high-end setups. The app's interface was designed to be as easy to use as possible for optimized integration into a caregiver's day-to-day responsibilities.

### 4 Evaluation

The studies reported here are part of the formative research examining *Healing Spaces* and should be considered in the context of a larger continuum of evaluation. Following conceptualization and early development, evaluation for usability and feasibility were conducted in two sequential steps. By reporting on these two early studies, we intend to emphasize the importance of the involvement of all human-computer interaction professionals in all stages of research and development for health-related products [13].

Throughout the development of our intervention, we documented design process and decision-making, which was critical for several reasons. Rigorous documentation of "existential" design decision-making is as valuable as empirical studies for the purpose of reproducibility, but also the development of theory-driven interventions [12]. On a practical note, the commercial success of most interventions for vulnerable populations is low and therefore preserving some of this effort at least as it was phenomenologically experienced is valuable for future efforts. Lastly, designing for and with vulnerable populations is always sensitive and difficult; therefore, evaluation, especially in early stages, should

## 4.1 Study I: Usability Evaluation

**Methods.** The RITE (Rapid Iterative Testing and Evaluation) method was led by the second author to ensure the app-based technological components of the experience were functional and easy to use. The RITE method is useful for identifying the most obvious usability problems using a small number of participants and implementing changes in a short amount of time [18].

**Participants.** Due to the need to quickly improve the user interface before the feasibility evaluation with the target users (older adults with dementia), the development team chose to conduct this study with participants from a convenience sample of students recruited at the University of Southern California. They were primarily graduate students in the Interactive Media and Games Division. RITE studies do not require IRB submission.

**Procedures.** For the purpose of the RITE study, only the iPad app and smart lights were used. Participants were given task-oriented instructions which included confirming that the lights were connected over WiFi and setting up and beginning an experience of their choice (see Figure 6). Testing sessions lasted between 10 and 15 minutes per participant. The app was updated twice during the RITE study to implement changes to address issues identified by participants.

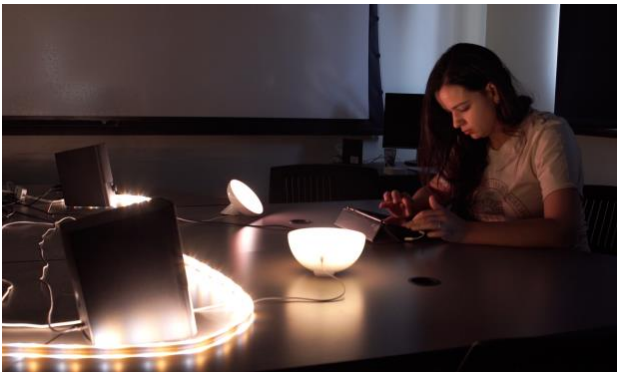

**Figure 6: Setup for the purpose of the RITE study. Image credits: Gabriela Purri R. Gomes.**

**Data Collection and Analysis.** Issues identified by participants were categorized by type: errors, failures, complaints, missed occurrences, and feature requests. *Errors* were anything a participant did wrong or any problem encountered that they could not easily resolve, but that were non-fatal to completing the task (e.g., trying to 'swipe' when a 'tap' gesture was required to open the Settings panel). *Failures* were issues that required intervention from the team for the participant to complete the task (e.g. problems with the WiFi connection). *Complaints* were issues of preference expressed by the participant (e.g., wanting more animations in the visual scene). A *missed* issue was noted if a participant did not notice something relevant to the task (e.g., the changes of the smart lights' colors). Lastly, *feature requests* were major suggestions for new features that were not currently planned for the app. Categorizing data this way helped the team prioritize which issues to address, with failure being the most important and feature requests being the least important to consider.

## 4.2 Study II: Feasibility Evaluation

**Methods.** A feasibility evaluation was then conducted at a memory care facility in order to investigate opportunities for improvement and usability of *Healing Spaces* with its target audience: older adults with dementia and their caregivers. By focusing on creating an "experiential space" where older adults with advanced dementia are able to express themselves through social, embodied interactions, this feasibility evaluation sought to elicit observable behavior to actively engage and recognize those with compromised cognition as co-creators of meaning and design. This study involved direct observation of *Healing Spaces* sessions and semi-structured interviews with caregivers and facility managers. The objective was to gain a better understanding of *Healing Spaces*' user experience in the context of a memory care facility and caregivers' overall satisfaction with a prototype involving only the most essential features.

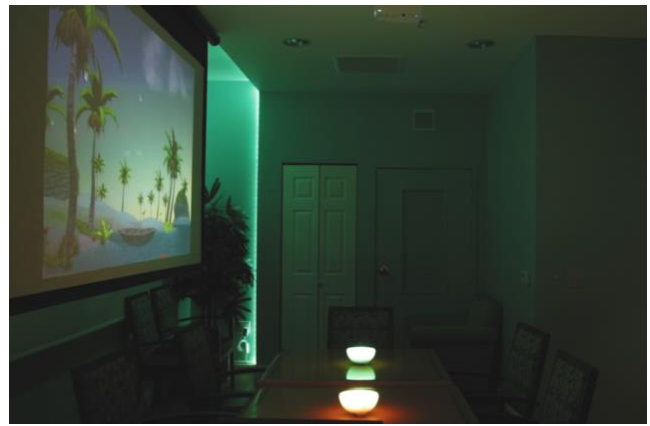

**Figure 7: The facility's sensory room with the Healing Spaces components in place. Image credits: Gabriela Purri R. Gomes.**

**Participants.** Staff of the facility identified eight residents with advanced dementia to recruit to participate in the study. These residents were already regularly utilizing the facility's sensory room with their caregivers. The project was submitted to the University of Southern California University Park Institutional Review Board under study ID IIR00002374 and activities were deemed exempt from requiring human subjects research review. The memory care facility followed their own procedures to obtain consent from all caregivers and responsible parties of the participating individuals with dementia. Five residents participated in *Healing Spaces* sessions twice over the course of the evaluation period, and the remaining three experienced it once. Six caregivers and two facility managers were involved in the observed sessions.

**Training.** Prior to the start of the feasibility evaluation, training sessions were held with participating caregivers at the facility. Training focused on teaching them how to operate each component of the system, including the iPad app, so that caregivers would feel comfortable using the system and leading a *Healing Spaces* session without assistance or interference. The purpose and scope of the *Healing Spaces* experience was also explained. Each training session lasted approximately 40-60 minutes and printed manuals with step-by-step system operation instructions were provided. Caregivers did not receive further instruction on how to conduct a session with *Healing Spaces* as the development team was interested in how experienced, professional caregivers would instinctively interact with the components presented; in this way, through this feasibility evaluation, the team was able to bring caregivers and participants informally into the design process to inform future iterations of *Healing Spaces.*

**Procedures.** The study took place at a memory care facility that serves older adults with various stages of dementia. The facility features a sensory room where a variety of activities involving sensory stimulation are provided; the space is set up with tables and chairs (see Figure 7). To accommodate the feasibility evaluation for *Healing Spaces*, a projector connected to an Apple® TV was installed in the room, and various types of smart lights were placed on tables or attached to the walls. Sound was delivered via a portable Bluetooth speaker.

Both individual (only one resident in the sensory room at a time) and group sessions (up to 3 residents at a time) were observed and recorded on video. Each session lasted 15 minutes. At least one representative from the development team was present alongside the caregiver:resident dyad, and sometimes other caregivers or facility managers would observe sessions. The feasibility evaluation lasted one week, with an average of two sessions taking place each day.

**Data Collection**. Direct observation of each session was conducted by the development team to uncover initial impressions of the overall user experience, to identify and address any negative reactions, and to gain first-hand insights into challenges faced by participants and potential opportunities for improvement. Additionally, semi-structured interviews were conducted with caregivers and facility managers after the sessions to discuss their overall impressions and thoughts on *Healing Spaces*. Finally, at the end of the evaluation week, a debrief session was held with the two facility managers to further discuss the perceived impact of the program on residents, caregivers and the facility as a whole. All sessions and interviews were recorded on video.

**Data Analysis.** Debrief meetings were regularly held at the end of each day with the development team and facility managers to reflect on field notes and to share findings, emerging insights and ideas. The video recordings of all sessions, interviews and debrief meetings were reviewed by the first author, and significant statements or quotes were noted and cross-referenced with field notes and insights that emerged from previous discussions. Findings were further refined with ongoing dialogue between the team and project advisors.

## 5 Results

### 5.1 Study I: Usability Evaluation

Overall, 29 unique issues were identified during the usability evaluation using the RITE method. The most frequent issue type identified was a complaint; failures were rare (see Table 1).

The team addressed a total of eight of the 29 identified issues via software updates. All eight (100%) of the issues that the development team agreed to address did not reoccur after the updates and were therefore considered 'fixed'. The primary reasons for not addressing the additional 21 issues were: issues of low severity or priority, disagreement on what changes to implement, requests being out-of-scope, or inappropriateness for the target end user or setting for the feasibility evaluation. This was an expected outcome for a RITE study.

**Table 1: Issue frequency by participant (P).**

|  | P1 | P2 | P3 | P4 | P5 | P6 | P7 | Totals |
|---|---|---|---|---|---|---|---|---|
| **Errors** | 1 | 1 | 1 | 2 | 1 | 0 | 1 | 7 |
| **Failures** | 0 | 1 | 1 | 1 | 0 | 0 | 0 | 3 |
| **Complaints** | 4 | 4 | 7 | 2 | 1 | 4 | 1 | 23 |
| **Missed** | 3 | 1 | 2 | 1 | 0 | 0 | 0 | 7 |
| **Feature Request** | 0 | 1 | 2 | 1 | 1 | 0 | 0 | 5 |
| **Totals** | 8 | 8 | 13 | 7 | 3 | 4 | 2 | 45 |

It is also worth noting that the remaining issues the team chose not to address did not necessarily reoccur, and that there was a downward trend in the total number of identified issues as the study progressed (see Table 1). This was likely due to general improvements to the app made by the development team during this time (not targeted at fixing an issue revealed by the RITE study participants) and/or, perhaps, idiosyncrasies of participants.

There were challenges to conducting the RITE study with non-target users in a setting without the full set-up. For instance, most participants were so engrossed in the iPad screen that they did not notice the smart lights changing colors during the experience, and many of the complaints documented were related to the lack of interactivity of the scene displayed on the iPad once they began the experience. This was likely due to a number of factors: that the scene was not projected onto a screen; that participants did not have a thorough understanding of what the final experience would be (no sensory box in this phase); and that participants were largely from an interactive media design program and thus expected a higher degree of digital interactivity which would not have been appropriate for the targeted end user. In contrast, during the feasibility evaluation it was found that caregivers tended to immediately put the iPad aside once the experience was started in order to focus on the resident and the sensory boxes. These differences did not impact the success of the RITE study overall, since they were not relevant to the main purpose of the study (identifying usability issues with the app component), and the design team had adequate knowledge about how the program would ultimately be used for the target population. However, using non-target users as participants may have impacted the efficiency of user testing by diverting time and attention toward less relevant issues and complaints and away from other issues that may have been identified. Despite these limitations, no severe system usability problems arose during the feasibility evaluation. Therefore, the team considered the RITE study to be a valuable step in the development and evaluation of *Healing Spaces,* successfully helping to identify and fix usability issues that would have otherwise impacted the feasibility evaluation.

## 5.2 Study II: Feasibility Evaluation

Overall, caregivers found the interface of the *Healing Spaces* app to be intuitive, simple, and easy to use. There were no noticeable struggles with the operation of the app during the pilot. There were some challenges related to the connectivity of all technological components with consequences for long-term maintenance as the digital components of the system are totally reliant on a working WiFi network. The setup involving an Apple® TV and projector often required troubleshooting by the observing team member during sessions. The experience could have been improved with better troubleshooting resources that could be easily accessible by the caregiver. Although the app was fast and easy to use, the set-up (e.g., lights, sensory boxes) was sometimes a challenge for the caregiver, especially if the room had not been cleaned up after utilization for another activity not related to *Healing Spaces*.

Although the same version of the app was used throughout the study, other aspects of the experience were adjusted in response to feedback and potential issues or situational factors as they arose (e.g., a portable light having to be moved away from participants because of the excess of brightness). At the end of each day, a debrief session was held with facility managers to discuss what could be improved for the next session. On the first day, for example, the team noticed that caregivers appeared to think they were expected to utilize every single element in the sensory box, and some residents did not respond well to that directive approach. The team then asked the facility's management to discuss changing this approach with the caregivers. Another practical insight was that the objects in the sensory boxes should be larger for easier handling and to discourage residents placing them in their mouths.

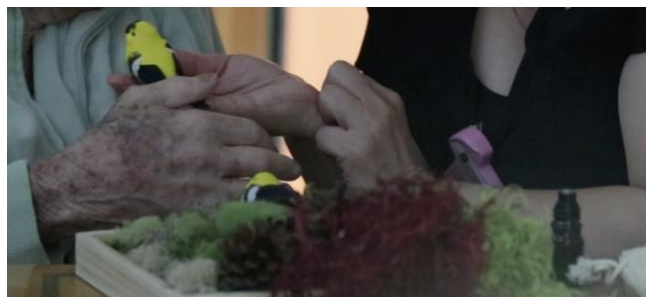

**Figure 8: Patient and caregiver interacting with the forest sensory box. Image credits: Gabriela Purri R. Gomes.**

**Qualitative Responses**. Overall, *Healing Spaces* provided opportunities for residents to engage their senses in all modalities (e.g., visuals, soundscapes, scents, and tactile). Caregivers reported that residents also appeared to be more aware of their surroundings and would look around the space. None of the sessions ended early and in one instance, a resident expressed not wanting to leave the experience by trying to keep her wheelchair at the table in the sensory room.

Observations of residents' emotional and cognitive states were frequently discussed by the caregivers during the interviews, although quantitative measurement of agitation, engagement, and/or alertness were not collected as part of the feasibility study. Most of the observed potential of *Healing Spaces,* as reported by caregivers, focused on the experience's impact on increasing engagement in residents. Both staff and caregivers expressed how impressed and surprised they were at how much some residents participated when using *Healing Spaces*. Residents were seen and described as *coming out of their isolation*, and *connecting with the environment or with the person next to them* more frequently than would be expected. As one caregiver explained in their interview,

> "It's a very interactive program (…) The experience I have just lived was incredible because the two residents that I have are sleeping all the time. Even if you talk to them, whatever we do, they don't stay awake. And it's amazing with these tools that you gave us … it's working [the residents are awake and engaged]."

Another caregiver described:

> "With our residents, especially those with cognitive impairment, it's hard for us to keep them awake or interested. I noticed with that with this project they try, they try to do something, even if they don't understand, they participate … and that's amazing."

Observations and comments such as these suggest that residents enjoyed the experience of *Healing Spaces*. When interviewed, some caregivers mentioned that the experience

was relaxing for themselves in addition to the resident. One caregiver remarked:

> "I think the relaxing environment is not only for them [residents], it's for us too. It's our time. When we are with a resident, it's time for them but it's also our time to feel more calm … so it's good."

In general, caregivers reported being satisfied with the experience of using *Healing Spaces* with residents. Multiple caregivers asked for more environmental options (e.g., a homely kitchen or a garage workshop). Though it was beyond the scope of the pilot to develop and implement additional environments, it suggests that caregivers were interested in the content of *Healing Spaces* and wanted to see more. Both caregivers and staff reported that they were interested in continuing to use *Healing Spaces* with the residents of the facility. Other comments from interviews describe aspects of *Healing Spaces*' design that were considered unique and appealing:

> "Normally, our sensory activities are done one sense at a time … whereas this [*Healing Spaces*] seems to be all enveloping where you're taking all the senses and creating an environment where they're getting the full experience. That is the most exciting, engaging part."
> (Facility manager)

> "A lot of our residents, because of the condition they suffer from, they don't like to get out. When they do, it's normally here in our garden. So bringing them here, to *Healing Spaces*, allowing them to see the lights, the color changes, the little boxes, the moss … I think it will be great for them. A lot of them don't get to feel the joy of traveling again … It's nice to know they don't have to travel, they can come here and it can be as close to traveling as they can get to."
> (Caregiver)

## 6  Discussion and Future Directions

Results overwhelmingly support the preliminary success and feasibility of *Healing Spaces.* The usability and feasibility studies helped the development team identify opportunities for technical and practical improvements as well as broader goals for the next iteration of *Healing Spaces*. The first improvement is to offer more customization in the platform for caregivers who have knowledge of and experience with the older adult receiving care (e.g., incorporating personally meaningful photos, objects or songs). A second improvement is to offer a more structured protocol for caregivers to follow due to the varying levels of caregiver skill in "improvising" how to use *Healing Spaces* to engage with the resident in the absence of detailed instructions. One possibility is to incorporate visual prompts into *Healing Spaces,* since visual changes in the virtual scene often captured caregivers' attention and sometimes sparked new behaviors for use of the sensory box and/or different conversation topics. However, future work will need to examine if the additional guidance impacts the overall experience, ensuring that the prompts are not invasive or disruptive. As part of developing a detailed protocol, it may be useful to create different experiences based on goals tailored to the individual residents (e.g., a reminiscing/memory task versus a sensory engagement task versus an activity to reduce agitation). A third improvement is to enhance the accessibility of *Healing Spaces* for older adults with a variety of abilities; for instance, many of the residents had significant hearing impairments and this was not adequately accounted for in the design of the auditory stimuli provided in the experience.

Next steps include a mixed methods examination of the efficacy of *Healing Spaces* to increase engagement and relaxation and reduce agitation in older adults with advanced dementia. This work will include both behavioral and physiological measures of agitation, engagement, and alertness, and will also examine possible mediating and/or moderating variables that impact the success of the intervention.

Additionally, interview data from the feasibility study suggested that *Healing Spaces* may have benefits for caregivers; therefore, the use of *Healing Spaces* to decrease caregiver burden and support caregiver wellness should be also investigated. According to the World Health Organization, there are currently an estimated 47 million people living with dementia worldwide, a number that is set to increase to 75 million by 2030 [30]. Caregivers enable millions of older adults with illness or disability to continue living in their homes and participate in the community [21]. Although caregivers often find this role rewarding, many experience negative effects on their own wellbeing, including economic burdens, stress, and contributions to declines in mental health [8] [29] [23]. Respite from caregiving can greatly improve caregiver wellbeing; however, respite care is often too expensive [25], and caregivers may feel discomfort leaving the person they assist with a respite provider [22]. Due to these challenges, technology and experiences like *Healing Spaces* offer a promising solution for caregivers struggling to obtain support.

*Healing Spaces* was showcased at festivals and exhibitions in the United States in 2018: the Electronic Entertainment Expo (E3), IndieCade (International Festival of Independent Games), and the University of Southern California's annual Games Expo. In all three showcases, the public reception was overwhelmingly positive. Attendees not only shared their own stories of struggle with loved ones who suffer from dementia, but also commented on the possible practical applications of *Healing Spaces* to populations other than adults with advanced dementia. Therefore, future research should also examine *Healing Spaces'* therapeutic value beyond dementia care, such as how it might be utilized to treat other types of sensory perception issues and behavioral disorders and potentially be utilized in settings such as hospitals, rehabilitation facilities and for children with disabilities in schools.

Ultimately, the development team aims to offer *Healing Spaces* as an "intervention in a box" -- a solution that presents all of its elements as an integrated whole that is easy to set up, use, and understand, with potential for expanded content.

**Limitations.** First, due to the flexibility required during the feasibility study, study activities were conducted in a somewhat uncontrolled environment (e.g., different furniture arrangements in the room). Second, due to the interest in *Healing Spaces,* facility managers and caregivers sometimes also observed sessions. The presence of these individuals and/or the research team members may have impacted the resident's experience of *Healing Spaces.* For example, the team documented that some residents would sometimes become distracted by other people in the room or the camera; their focus would need to be brought back to the various elements of *Healing Spaces* by verbal prompting. Lastly, as training provided to caregivers was purposefully general, without a strict protocol to follow, caregivers exhibited varying levels of skill in utilizing *Healing Spaces* to engage with residents in the absence of detailed instructions.

## ACKNOWLEDGEMENTS

The first author would like to thank her project team as well as her advisors and mentors Dennis Wixon, Margaret Moser and Nina C. Poulos Ramage. We'd also like to thank the residents, caregivers and staff for their time and participation in the research and development of *Healing Spaces*. This unfunded project was completed as part of the first author's MFA thesis work at the University of Southern California. Dr. Stein Duker was supported by the National Institutes of Health under NCMRR K12 HD055929.